\documentclass[conference]{IEEEtran}
\IEEEoverridecommandlockouts

\usepackage{cite}
\usepackage{mathtools}
\usepackage{enumerate}
\usepackage{braket}
\usepackage{amsmath,amssymb,amsfonts}
\usepackage{algorithmic}
\usepackage{graphicx}
\usepackage{textcomp}
\usepackage{xcolor}
\usepackage{hyperref}
\usepackage{subfig}

\begin{document}

\title{Evaluation of Parameterized Quantum Circuits with Cross-Resonance Pulse-Driven Entanglers\thanks{This work was supported in part by the QISE-NET NSF Fellowship DMR 17-47426}}
\author{\IEEEauthorblockN{Mohannad Ibrahim\IEEEauthorrefmark{1}, Hamed Mohammadbagherpoor\IEEEauthorrefmark{2}, Cynthia Rios\IEEEauthorrefmark{3}, Nicholas T. Bronn\IEEEauthorrefmark{4} and Gregory T. Byrd\IEEEauthorrefmark{5}}
\IEEEauthorblockA{\IEEEauthorrefmark{1}\IEEEauthorrefmark{2}\IEEEauthorrefmark{3}\IEEEauthorrefmark{5}\textit{Department of Electrical and Computer Engineering, North Carolina State University, Raleigh, North Carolina} \\
\IEEEauthorrefmark{4}\textit{IBM Quantum, IBM T.J. Watson Research Center, Yorktown Heights, New York, USA} \\
mmibrah2@ncsu.edu\IEEEauthorrefmark{1}, hmohamm2@ncsu.edu\IEEEauthorrefmark{2},
cvrios@ncsu.edu\IEEEauthorrefmark{3}
ntbronn@us.ibm.com\IEEEauthorrefmark{4}
gbyrd@ncsu.edu\IEEEauthorrefmark{5}}
}
\maketitle

\begin{abstract}
Variational Quantum Algorithms (VQAs) have emerged as a powerful class of algorithms that is highly suitable for noisy quantum devices. Therefore, investigating their design has become key in quantum computing research. Previous works have shown that choosing an effective parameterized quantum circuit (PQC) or ansatz for a VQA is crucial to its overall performance, especially on near-term devices. In this paper, we utilize pulse-level access to quantum machines, our understanding of their two-qubit interactions, and, more importantly, our knowledge of VQAs, to customize the design of two-qubit entanglers. Our analysis shows that utilizing customized pulse gates for ansatze reduces state preparation times by more than half, maintains expressibility relative to standard ansatze, and produces PQCs that are more trainable through local cost function analysis. Our algorithm performance results show that in three cases, our PQC configuration outperforms the base implementation. Experiments using IBM Quantum hardware demonstrate that our pulse-based PQC configurations are more capable of solving MaxCut and Chemistry problems compared to a standard configuration.
\end{abstract}


\section{Introduction}
\label{sec:introduction}
Looking at various limitations in current noisy quantum hardware, one might first think that the development of such systems at this stage relies heavily on quantum hardware engineers and experimental physicists. However, algorithm designers have successfully contributed to pushing limitations such as limited numbers of qubits, limited qubit connectivity, and coherence times by designing algorithms tailored for such systems. For example, the Variational Quantum Algorithm (VQA) employs a quantum-classical approach to counter current device limitations. 

The general framework of VQA begins with identifying a problem-specific cost function. Next, a trainable Parameterized Quantum Circuit (PQC) or ansatz is used to evaluate this cost. This PQC is then trained in a hybrid quantum-classical loop that tries to minimize the cost. By pushing the parameter optimization load to the classical optimizer, VQAs are able to run short-depth circuits and hence are very suitable for current machines \cite{vqas}.

Two of the most prominent examples of VQAs are the Variational Quantum Eigensolver (VQE) \cite{vqe}, and the Quantum Approximate Optimization Algorithm (QAOA) \cite{qaoa}. With VQAs being one of the most promising candidates for demonstrating advantage, and with various companies and institutes releasing devices with $10$s--$100$s of qubits, VQAs have become one of the most investigated topics in quantum computing research. They are established as major quantum workloads, with researchers proposing optimizations to their implementation through all layers of the quantum computing stack \cite{vqas}.

In this work, we demonstrate how a deeper understanding of quantum device control can impact VQAs. We optimize VQAs by exploring the lowest level of quantum control: pulse-level access \cite{qiskit_pulse,qiskit_pulse_2}, targeting the most integral part of the algorithm, its ansatz. There exist several ansatz architectures: ones that are \textit{problem-specific} and others that are \textit{problem-agnostic} creating Hardware Efficient Ansatz (HEA) \cite{kandala}. We focus mainly on HEA in this paper.

One major problem VQAs encounter is the occurrence of barren plateaus in PQC training. It has been proven that if an anstaz is sufficiently random, the gradient of the cost function vanishes exponentially with the number of qubits~\cite{mcclean_bp}. Therefore, the majority of studies on PQCs~\cite{mcclean_bp, cerezo_bp, patti_bp, ent_bp, how_much_ent, grant_init, meta_vqe, q_gradient, layer_pqc, aavqe, sukin_pect, axis_vqe} focus mainly on trainability and optimization procedures, and lesser attention is given to their device-specific performance and optimization. More recently, hardware-oriented analysis and optimization of PQCs has been explored in \cite{vqe_noise, vqe_extra_2, vqe_extra, vqe_ravi, vqe_pulse, avqe, ctrlVQE, vqe_pulse_2, evqe, noise_bp}. A core design challenge in this direction is realizing the degree to which hardware should influence the algorithm implementation without sacrificing performance~\cite{mcclean_bp}. Thus, our goal in this paper is to identify a suitable combination of algorithm and hardware metrics that can guide pulse-level VQA optimization approaches.

From the hardware side, we utilize Hamiltonian Tomography (HT) \cite{sheldon_cr}, an accurate Hamiltonian calibration technique, to characterize our pulse implementations. We utilize HT to benchmark and customize the cross resonance (CR) gate \cite{sheldon_cr, chow_cr, stancil_byrd}, the entangling gate used by IBM's superconducting backends. We utilize the results obtained from HT to analyze our PQCs for the algorithmic descriptors: expressibility  \cite{sukin_expr}, entanglement entropy, and trainability \cite{mcclean_bp, cerezo_bp, patti_bp}. Our pulse-driven PQCs achieve a speedup of up to 2.9x with an average of 2.51x over a base PQC design. We demonstrate VQE performance for MaxCut and Chemistry benchmarks on IBM's $27$-qubit machine \textit{ibmq\_montreal}, accessible through the IBM Quantum cloud service. Our algorithm performance results show that in at least three cases, the pulse-driven PQC configurations outperform the base PQC in trainability and solution quality.
\begin{figure}
    \centering
    \includegraphics[width=3in, keepaspectratio]{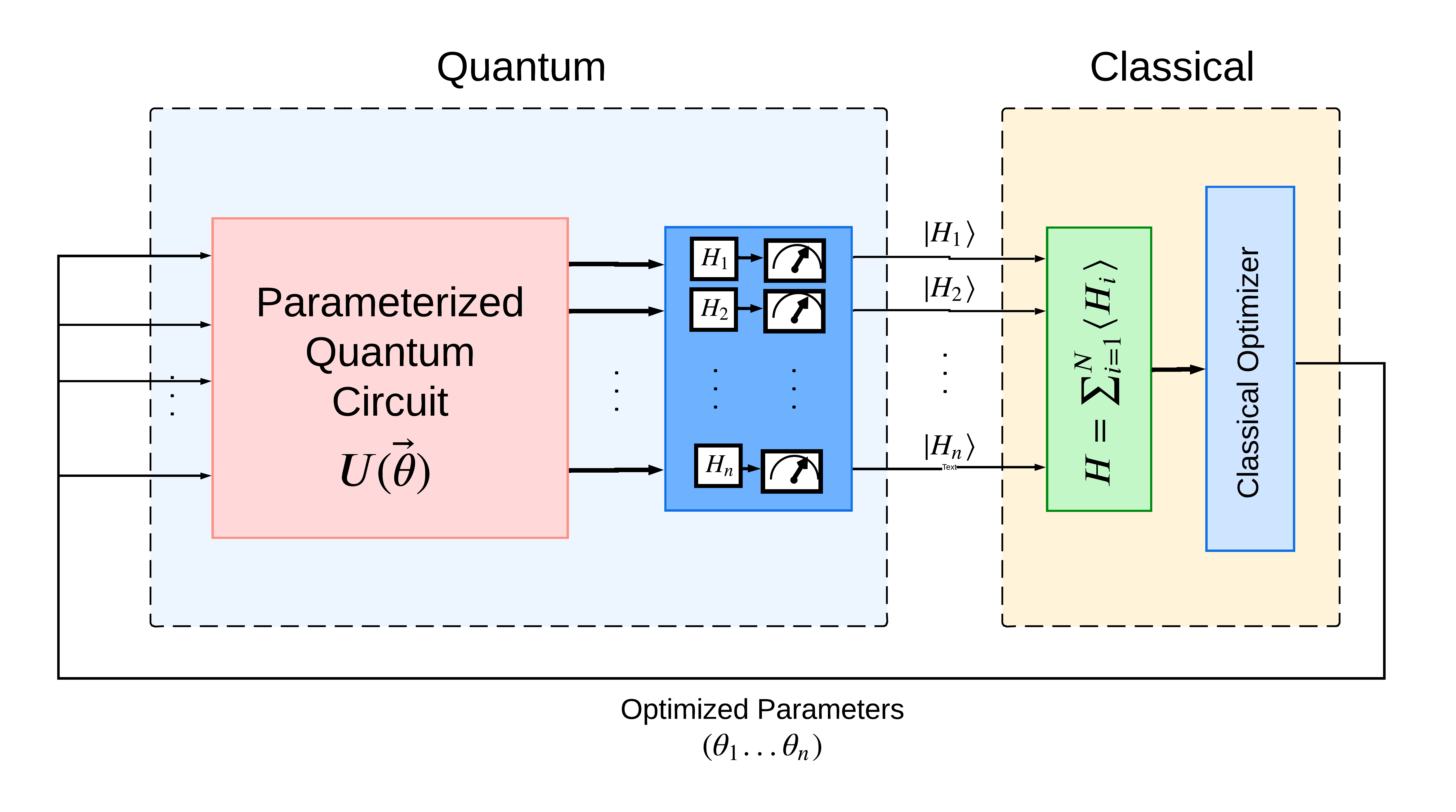}
    \caption{Block Diagram of the VQE Algorithm.}
    \label{fig:VQE}
\end{figure}
\section{Background}
\label{background}
\subsection{Quantum Computing Basics}
% A \textit{qubit} is the fundamental unit of information in quantum computing which is analogous to a bit in classical computing. While a classical bit can be in one of two states $\{0,1\}$, a qubit can be in a linear \textit{superposition} of those two, i.e., $\ket{\psi} = \alpha\ket{0} + \beta \ket{1}$ where $|\alpha|^2+|\beta|^2 =1$ and $\alpha, \beta \in \mathbb{C}$. The states $\ket{0}$ and $\ket{1}$ are the computational basis states.
One of the essential differences between classical and quantum computing algorithms is that the latter are intrinsically probabilistic models. The quantum \textit{measurement} operation collapses the ``definite'' state of a qubit in a two-dimensional \textit{Hilbert} space to one of the two computational basis states. \textit{Quantum operations} or \textit{gates} are used to manipulate/modify information stored in qubits. A gate is defined by a unitary operation that can be considered as a rotation over the \textit{Bloch Sphere}~\cite{mike_n_ike} and can either act on single or multiple qubits. The physical implementation of gates depends largely on the type of quantum hardware. For example, in IBM superconducting quantum computers, microwave voltage pulses are applied to qubits \cite{murali_xtalk,z_gate} to implement the gates. The same principle is used to implement \textit{entanglement} between qubits, which results in non-classical correlated effects~\cite{murali_xtalk}. 

Today's implementations of quantum workloads are constrained by limitations in current noisy quantum hardware. In the past decade, tremendous efforts have been made to improve the fidelity of quantum hardware, along with algorithms specifically targeting current and near-term machines. A major class of such algorithms is the variational quantum algorithm (VQA).

\subsection{Variational Quantum Algorithms}
A \textit{Variational Quantum Algorithm (VQA)} is a hybrid scheme of computation that allocates tasks to both quantum and classical computing resources and coordinates the execution between the two through a tight feedback loop to achieve a larger computational goal. In contrast to quantum algorithms developed for the fault-tolerant era, VQAs are highly suitable for current noisy quantum hardware. This suitability stems from utilizing classical optimizers for parameter tuning, which helps keep the quantum circuit depth shallow, hence mitigating noise. 

The algorithm's modular structure and suitability to current and near-term systems have led to its widespread use. In fact, exploring various aspects of VQAs is a key part of the research on quantum systems, and identifying the conditions under which this class of algorithms will succeed is still an open question \cite{preskil_nisq}. VQAs have been applied to a wide variety of applications \cite{vqas} such as quantum chemistry \cite{kandala, chem_1,chem_2,chem_3}, combinatorial optimization \cite{comb_1,comb_2,comb_3}, and machine learning \cite{expr_5,entang_3, ml_1, ml_2, expr_4}. A complete discussion of VQAs can be found in the review paper by Cerezo \textit{et al.}~\cite{vqas}. 

A prime example of VQAs is the Variational Quantum Eigensolver (VQE) \cite{vqe} shown in Fig. \ref{fig:VQE}. The trial wave function $\psi_\theta$ is generated by applying the PQC ($U(\vec{\theta})$), which is expected to explore the Hilbert space efficiently. Once the trial state is prepared, the expectation value of the problem Hamiltonian $H$ is determined. The Hamiltonian first needs to be decomposed or ``mapped'' from its original form (e.g., fermionic modes) to spin (Pauli) operators in a way that preserves the commutation relations~\cite{soma_2002}. Once decomposed, $H$ can be represented as $H=\sum{a_iP_i}$, where Pauli string $P_i$ is the tensor product of Pauli operators.

VQE utilizes classical optimization to find suitable parameters for the PQC, with the goal of minimizing the expectation value of $H$. The variational principle guarantees that the expectation value $\langle H \rangle$ is always greater than the minimum eigenvalue of the system (the ground state energy $E_0$). The classical optimizer is applied iteratively to update the PQC parameter set ($\vec{\theta}$), and a quantum computer is used to compute information about the Hamiltonian's expectation value for the calculated $\vec{\theta}$ based on the measurements. The algorithm is repeated until convergence, or an optimizer limit is reached. Various types of optimization procedures, such as gradient descent algorithms or direct search methods, can be used to update the circuit parameters \cite{vqas}.
\subsection{Parameterized Quantum Circuits}
A \textit{Parameterized Quantum Circuit (PQC)} or \textit{ansatz} is defined as a tunable unitary operation $U(\vec{\theta})$ that is applied to a quantum state $\ket{\psi_0}$, often initialized to $\ket{0}^{\otimes n}$ \cite{sukin_expr} or a problem-influenced initial state. This results in the quantum state
\begin{equation}
\ket{\psi_\theta} = U(\vec{\theta})\ket{\psi_0}
\end{equation}
where $\vec{\theta}$ is a vector of a polynomial number of circuit parameters. These parameters can represent any tunable feature of a quantum operation, but they usually correspond to angles of rotation gates. A PQC can be further decomposed to a product of $L$ sequentially applied sub-unitaries \cite{vqas}, usually referred to as  \textit{layers}
\begin{equation}
U(\vec{\theta}) = U_L(\vec{\theta_L}) ... U_2(\vec{\theta_2})U_1(\vec{\theta_1})
\end{equation}
This modular nature of PQCs has been recently compared to classical computing, in which the parameters of the PQC are analogous to the weights and biases of a classical neural network \cite{sukin_expr}. Similar to the broad spectrum of neural network architectures, PQC designs can vary widely in their design goals and performance. Nonetheless, they can be classified into two main types: a \textit{problem-specific} approach that utilizes knowledge of the problem to tailor the PQC architecture \cite{vqe_noise, pqc_1, pqc_2}, and a \textit{problem-agnostic} or \textit{hardware-efficient} design \cite{kandala} that focuses more on the suitability of the design to hardware \cite{vqas}. This paper focuses mainly on the latter type, hardware-efficient PQCs, and expands on various aspects of their design.

\begin{figure}[t]
  \begin{center}
  \includegraphics[width=3in]{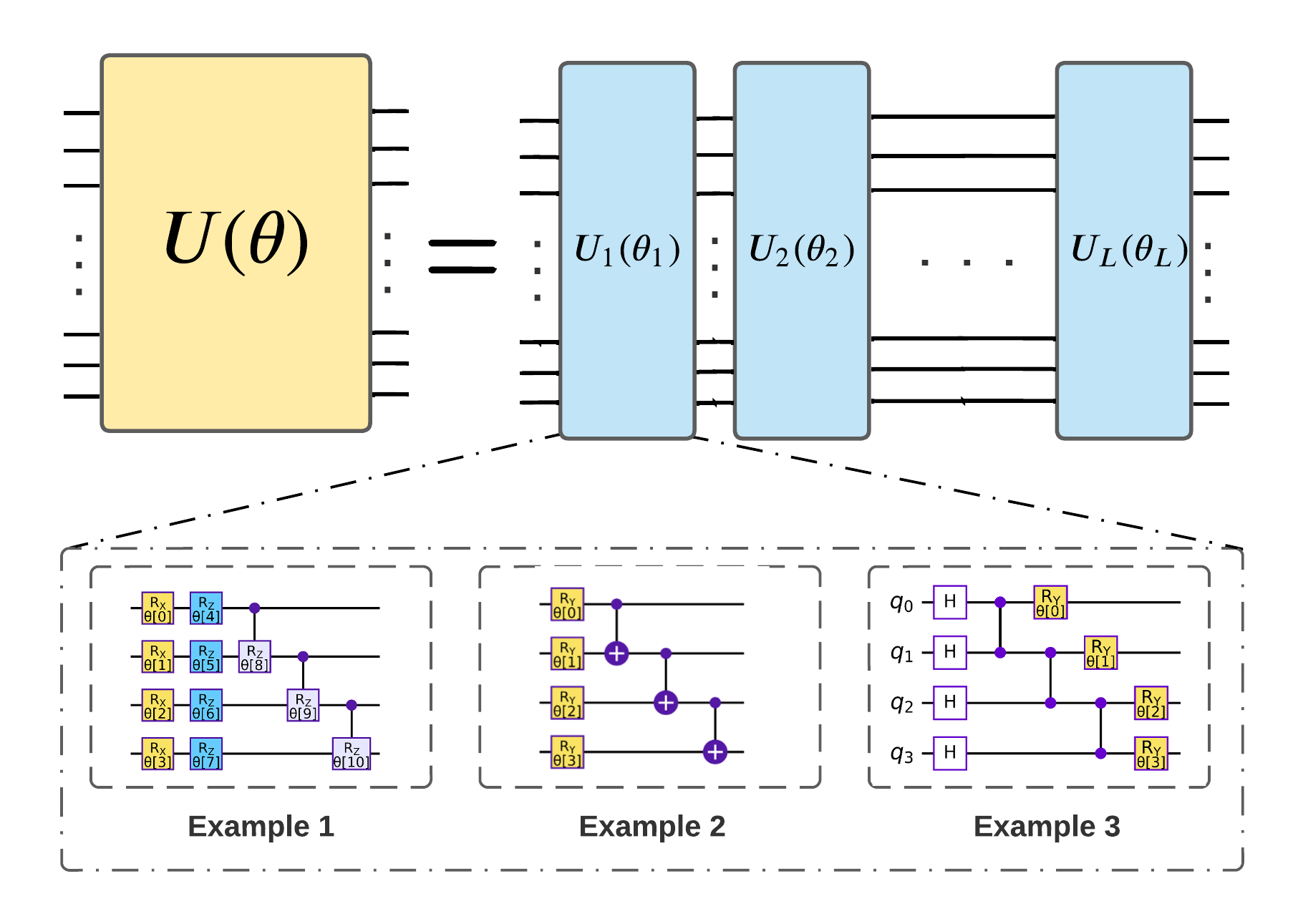} \\
  \caption{Schematic diagram of a Hardware-Efficient PQC with example layer designs.}
  \label{fig:PQC}
  \end{center}
 \vspace{-2em}
\end{figure}

Hardware-efficient PQCs generally aim at reducing both gate count and circuit depth. A single layer in this approach is usually composed of single-qubit operations followed by entangling two-qubit operations based on the physical connections of the hardware. A multi-layer PQC with this approach has been shown to be more suitable to current noisy machines compared to unitary coupled-cluster ansatze \cite{kandala}. Fig. \ref{fig:PQC} shows examples of layer designs following this strategy. Additionally, other device-specific information such as gate decomposition, physical connections between qubits, crosstalk, and other noise characteristics can also influence the design choices for hardware-efficient PQCs.

Besides their structure, classifying and understanding the usefulness of different PQC designs is necessary for better VQA design. The next section (\ref{expr_entang}) expands further on the topic.

\subsection{Expressibility, Trainability, and Entanglement}
\label{expr_entang}
With the wide range of PQC architectures, a fundamental question is whether a circuit can adequately prepare the target quantum state. In this regard, researchers have proposed different metrics to estimate the quality of an ansatz \cite{sukin_expr, expr_2, expr_3, expr_4, expr_5, holmes_expr}. In this section, we describe the three qualitative metrics we used in this paper to estimate a PQCs \textit{expressibility}, \textit{trainability}, and \textit{entanglement}.
\subsubsection{Expressibility}
Proposed by Sim \textit{et al.} \cite{sukin_expr}, Expressibility (Expr) is defined as a PQC's ability to produce quantum states that well represent the Hilbert space. The general idea is to compare the distribution of states obtained from a PQC's $U(\vec{\theta})$ to the maximally expressive uniform \textit{(Haar)} random states. By sampling pairs of parameter values and their associated quantum states (e.g. using $U(\vec{\theta_1})$ and $U(\vec{\theta_2})$), we can compute the probability distribution of the quantum state fidelities $\hat{P}_{\text{PQC}}(F;\vec{\theta})$ \cite{sukin_expr}. Expr is then estimated using the \textit{Kullback-Leibler} divergence $D_{\rm KL}$ \cite{kl} as follows
\begin{equation}
{\rm Expr} = D_{\text{KL}|}(\hat{P}_{\text{PQC}}(F;\vec{\theta}) || P_{\text{ Haar}}(F))
\end{equation}
where $P_{\text{ Haar}}(F)$ is the probability distribution of fidelities for the Haar random state. Please refer to the paper~\cite{sukin_expr} for more details on this metric. In short, a smaller Expr value for a PQC indicates a closer approximation to random states and, hence, a more expressible circuit.

\subsubsection{Trainability}
A more expressive ansatz does not necessarily lead to better VQA performance. It is also essential to characterize the properties of the VQA's optimization landscape and employ efficient training routines to guarantee performance. Perfectly expressive ansatze are actually proven to have flatter optimization landscapes and thus are less trainable \cite{holmes_expr}. The first work to investigate the trainability of PQCs was by McClean \textit{et al.} \cite{mcclean_bp}. Their work proved that a wide variety of PQCs, particularly hardware-efficient ones, suffer from vanishing gradients exponentially in the number of qubits - a phenomenon known as \textit{barren plateaus}. This observation has been further expanded by Cerezo \textit{et al.} \cite{cerezo_bp} to indicate that the occurrence of barren plateaus is cost-function-dependent for shallow ansatze. In recent years, more works have shown that other factors can also impact barren plateaus, such as noise \cite{noise_bp} and entanglement \cite{ent_bp, patti_bp}.
% \begin{figure}[t]
%     \centering
%     \includegraphics[width=3in]{figures/circuit_pulse.png}
%     \caption{(a) A quantum circuit that prepares and measures a Bell state. (b) The pulse sequence that implements the circuit in (a) on \textit{ibmq\_montreal}. Channel $u_0$ is the control channel dedicated for two-qubit interactions between qubits $q_0$ and $q_1$. Note here that pulses are scheduled with an \textit{As-Late-as-Possible (ALAP)} method, which minimizes the idle time between instructions on the same channel and maximizes the qubit idle time before the first pulse \cite{qiskit_pulse}. The pulse envelopes are filled with bright and dark colors representing the real (in-phase) and imaginary (quadrature-phase) components of the waveform, respectively.}
%     \label{fig:circuit_pulse}
%     \vspace{-1em}
% \end{figure}

A general definition of a cost function can be
\begin{equation}
C = \bra{\psi}U^\dagger(\vec{\theta})\hat{O}(\omega)U(\vec{\theta})\ket{\psi}
\end{equation}
where $U(\vec{\theta})$ is the ansatz unitary acting on state $\ket{\psi}$, and $\hat{O}(\omega) = \sum_i \omega_i\hat{O}_i$ is an observable or Hamiltonian that acts nontrivially on a subset of the circuit qubits (\textit{local}) or the total circuit qubits (\textit{global}). The classical learning algorithm minimizes $C$ by updating a parameter $\theta_i$ through the use of the partial derivatives (i.e., $\frac{\partial C}{\partial \theta_i}$) which represents the contribution to the gradient $\partial C$ from the change in parameter $\partial\theta_i$. In Section~\ref{results}, we utilize cost-function-dependent barren plateau analysis \cite{cerezo_bp} to evaluate our PQC's trainability. We use $Var[\partial_iC]$, which represents the variance of the partial derivative of the cost function $C$ with respect to $\theta_i$ for $n$ sampled circuits. The magnitude of the variance quantifies the partial derivative's concentration around zero~\cite{cerezo_bp}. Thus, smaller values indicate less trainability.

\subsubsection{Entanglement}
Entanglement measurement quantifies the amount of entanglement contained in a quantum state. It is first essential to realize the advantages of generating highly-entangled states for VQAs. Prior works have shown that highly-entangled PQCs are potentially more capable of capturing non-trivial correlations in the quantum data and efficiently represent the solution space for tasks like the ground state preparation or data classification \cite{sukin_expr, entang_1, kandala, entang_2, entang_3}. On the other hand, excessive entanglement can possibly lead to concentration of measure, making a PQC too random and less trainable \cite{ent_bp}. In recent works, entanglement has been investigated as a primary source of barren plateaus \cite{ent_bp, patti_bp}. With such tradeoffs between entanglement and trainability, optimization problems vary in their utilization of entanglement for performance \cite{how_much_ent, ent_extra_1, ent_extra_2, ent_extra_3}. This ultimately leads to the importance of developing a comprehensive understanding of the role of entanglement in VQAs.

There exist several methods for quantifying entanglement \cite{sukin_expr, entang_meas_1, entang_meas_2, entang_meas_3}. In this paper, we use the bipartite entanglement entropy, which is the Von Neumann entropy of the reduced density matrix of any of the subsystems, to estimate the spread $S$ of circuit entanglement
\begin{equation}
\label{eqn:entropy}
S = Tr[\rho_\alpha\log_2\rho_\alpha]
\end{equation}
where $\rho_\alpha$  is the reduced density matrix of $(n-1)/2$ connected qubits containing as many cost function qubits as possible \cite{patti_bp}. In Section~\ref{results}, we analyze the entanglement of our PQCs by observing both $S$ and HT results and tie this to their trainability with respect to cost function size and number of layers.

\subsection{Pulse-Level Control of Quantum Systems}
The lowest level of control of a quantum computer is through \textit{pulse}s. Such a level of control can be realized/enabled by a classical microprocessor with an embedded pulse digital-to-analog converter \cite{qiskit_pulse}.
A \textit{pulse} is defined as a time-series of complex-valued analog amplitudes, each called a sample, applied to qubits on any type of input \textit{channels}, at each system cycle time \texttt{dt}~\cite{qiskit_pulse}. Typically, the timing of scheduled operations is inconsequential in the standard quantum circuit model as long as the order of non-commuting operators is preserved \cite{metodi_shed, qiskit_pulse}. However, timing considerations are very critical once we move to the pulse model on quantum hardware such as transmons~\cite{qiskit_pulse, qiskit_pulse_2}.

Quantum computers are routinely calibrated to account for drifts in their state by updating their experimental parameter settings~\cite{pranav_op,murali_xtalk, tannu_qubits}. Such calibrations are key to obtaining the translations from gates to pulses or pulse schedules. For example, current IBM quantum backends implement the $X$ gate as an almost-Gaussian \textit{DRAG} pulse \cite{drag} (with a carrier frequency equal to that of the ground-to-excited state transition), while $Z$ and $R_Z(\theta)$ gates are purely implemented in software~\cite{pranav_op}. Figures~\ref{fig:cr_cnot}(a) and (b) show a quantum circuit and its corresponding pulse schedule. We demonstrate pulse control of quantum systems using IBM's framework for pulse-level access, Qiskit Pulse \cite{qiskit_pulse, qiskit_pulse_2}. Table~\ref{table:qiskit_pulse} shows a summary of the different pulse channels used in IBM machines and their descriptions.

\begin{table*}[ht]
\centering
\caption{Qiskit Pulse Channels Summary}
\label{table:qiskit_pulse}
\resizebox{\textwidth}{!}{%
\begin{tabular}{|l|l|}
\hline
\textbf{Channel} & {\centering \textbf{Description}} \\ \hline
DriveChannel $d_i$ & Main drive channel connected to qubit $i$, with signals modulated at the resonance frequency of the qubit \\ \hline
MeasureChannel $m_i$ & Connected to the readout component of qubit $i$ \\ \hline
ControlChannel $u_i$ & Transmit channel associated with arbitrary interaction between specific 2 qubits $j$ and $k$ \\ \hline
AcquireChannel $a_i$ & Connected to the readout component of qubit $i$ to digitize and acquire measurement data \\ \hline
\end{tabular}
}
\end{table*}
 
\section{Methodology}
\subsection{Dissecting the Cross Resonance Gate}
\label{cr_dissect}
The \textit{Cross Resonance (CR)} gate is an all-microwave entangling gate, obviating the need for tunable qubits or couplers~\cite{sheldon_cr, rigetti_cr, chow_cr, stancil_byrd}. This feature makes for better scaling to larger numbers of qubits by minimizing the overhead of control electronics and control wires~\cite{rent, nereeja_cr}. Thus, the CR gate emerged as a promising two-qubit entangling gate in quantum architectures based on planar, fixed-frequency superconducting \textit{transmons} \cite{nereeja_cr, sheldon_cr}. Transmon qubits are designed to have reduced sensitivity to charge noise while maintaining sufficient anharmonicity, allowing the lowest two levels to be addressed as a qubit \cite{transmon}.

For a pair of coupled fixed-frequency transmons, a CR interaction is realized by driving the \textit{control} transmon at the frequency of the \textit{target} transmon. This interaction produces an effective Hamiltonian of the form \cite{qiskit_pulse}
\begin{equation}
\label{eqn:cr_ham}
\begin{aligned}
    \bar{H}_{CR} = & \ \frac{Z \otimes A}{2} + \frac{I \otimes B}{2} \\
    A = & \ \omega_{ZI}I + \omega_{ZX}X + \omega_{ZY}Y + \omega_{ZZ}Z\\
    B = & \ \omega_{IX}X + \omega_{IY}Y + \omega_{IZ}Z
\end{aligned}
\end{equation}
where each term represents Pauli operators applied to both control and target, with the control being first and the target being second in the tensor product reading from left to right. For example, the term $\omega_{IZ}\frac{IZ}{2}$ corresponds to Pauli-$I$ and $Z$ operators applied to the driven control and target qubits, respectively, generating an uncontrolled (because of the Pauli-$I$ on the control) $Z$-rotation on the target transmon of strength $\omega_{IZ}$.

\begin{figure}
    \centering
    \includegraphics[width=3in]{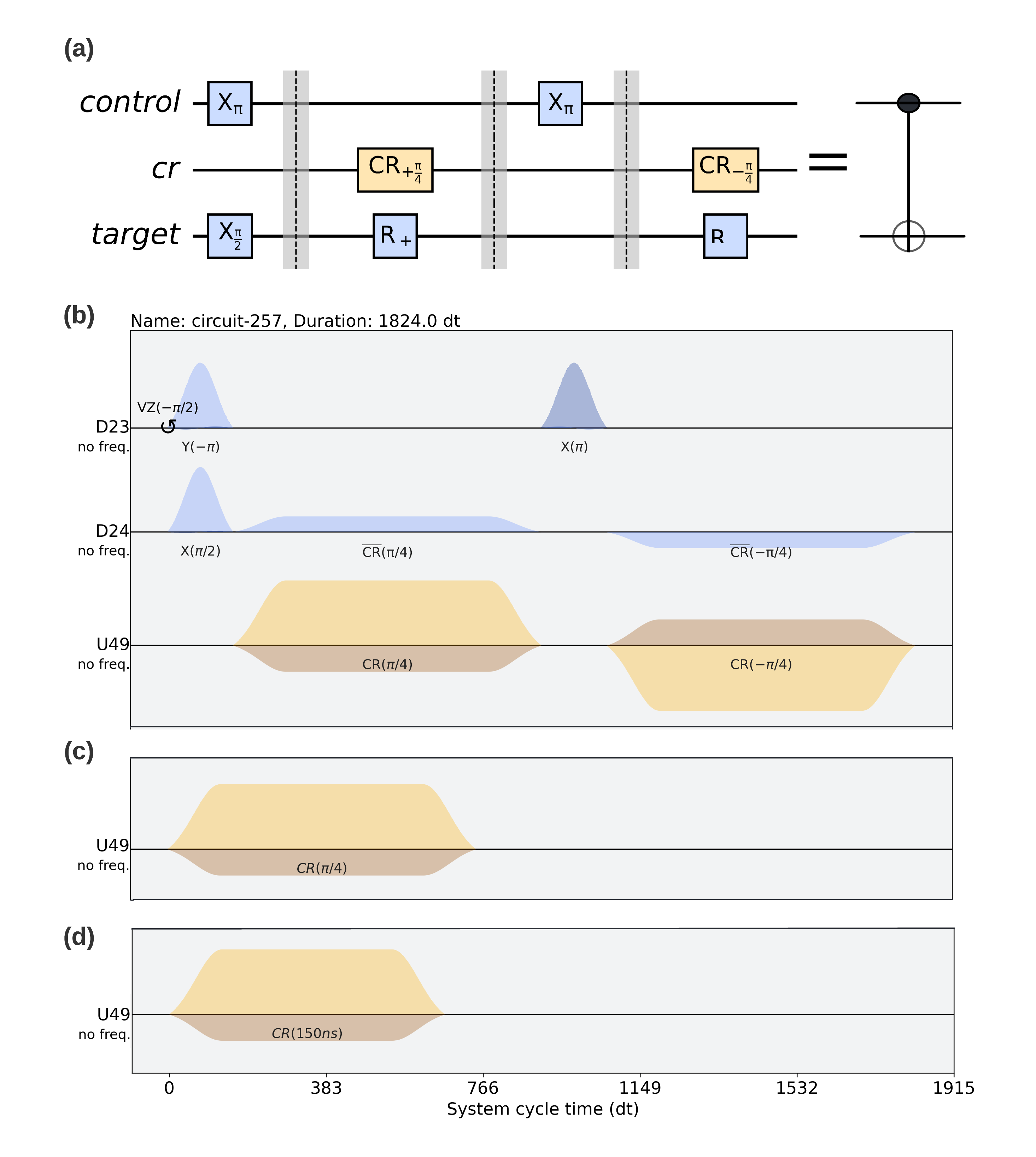}
    \caption{\textbf{(a)} The gate representation of the CR-based CNOT implementation. \textbf{(b)} The equivalent pulse sequence on \textit{ibmq\_montreal}. Note that all pulse parameters (durations, amplitudes, etc) shown are specific to qubits 23 and 24 from the device calibrations. The circular arrows represent phase shifts. Note here that pulses are scheduled with an \textit{As-Late-as-Possible (ALAP)} method, which minimizes the idle time between instructions on the same channel and maximizes the qubit idle time before the first pulse \cite{qiskit_pulse}. The pulse envelopes are filled with bright and dark colors representing the real (in-phase) and imaginary (quadrature-phase) components of the waveform, respectively. \textbf{(c)} The custom pulse implementation of $CR(\pi/4)$ for the same set of qubits. The pulse parameters are derived from the calibrations returned by the \texttt{instruction\_schedule\_map}. \textbf{(d)} The custom pulse implementation of $CR(150$~ns$)$, which uses a fixed duration of 150 ns.}
    \label{fig:cr_cnot}
\end{figure}

If isolated, the $ZX$ conditional rotation term in (\ref{eqn:cr_ham}), with rotation angle $\frac{\pi}{2}$, would result in the unitary $U_{ZX}(\frac{\pi}{2}) = e^{-i\frac{\pi}{4}ZX}$, which is locally-equivalent to the standard CNOT (i.e., equivalent up to single-qubit gates). The CNOT gate is sufficient for universal quantum computation when combined with arbitrary single-qubit operations \cite{nereeja_cr, qiskit_pulse}. However, the other terms in the effective Hamiltonian (\ref{eqn:cr_ham}) are coherent error terms and ``unwanted'' for generating the unitary equivalent to CNOT. Developing strategies to characterize and control these terms in order to create high-fidelity entangling gates is still ongoing research \cite{nereeja_cr, qiskit_pulse}. Figures \ref{fig:cr_cnot}(a) and \ref{fig:cr_cnot}(b) show IBM Quantum's standard echoed technique used to suppress these terms and implement the CNOT gate. This pulse sequence is comprised of three main components:
\begin{itemize}
\item \textbf{Echoed CR Pulses}: two CR pulses with opposite phases on the control channel $(u)$ and two single-qubit pulses on the drive channel, one before each CR pulse. This sequence (grouped with the echoed CR pulses) refocuses/echoes away unwanted terms (mainly $IX$ and $ZI$) in the interaction Hamiltonian \cite{corcoles_cr, qiskit_pulse}.
\item \textbf{Compensation Pulses}:
also known as \textit{target rotaries}, these are used to address and reduce errors identified in the echoed CR Hamiltonian arising from driven $ZZ$ interactions and classical crosstalk ($IY$). Additionally, they suppress unwanted entanglements with target nearest-neighbors or \textit{spectators} due to static coupling, without increasing the CR pulse length~\cite{nereeja_cr}.
\item \textbf{Single Qubit Pulses}: Additional single-qubit pulses on the control and target are used to build a CNOT from the generated $U_{ZX}(\pi/2)$ unitary. This sequence can be ``reversed'' in that the physical CR control qubit may be a logical CNOT target qubit with the appropriate addition of single-qubit gates.
\end{itemize}
In the next section (Section-\ref{custom_gate}), we use our understanding of the CR Hamiltonian and pulse sequence to implement a pulse-efficient entanglement gate suitable to VQAs.

\begin{table}[t]
\centering
\caption{Decomposition of Some IBM two-Qubit Gates \\ using Qiskit's Transpiler}
\label{table:2_qubit_gates}
\resizebox{\columnwidth}{!}{%
\begin{tabular}{|l|l|}
\hline
\textbf{Gate}                 & \textbf{Decomposition}        \\ \hline
Controlled-$Z$                  & 2 single-qubit gates, 1 CNOT  \\ \hline
Controlled-$R_y$                 & 2 single-qubit gates, 2 CNOTs \\ \hline
Controlled-Phase              & 3 single-qubit gates, 2 CNOTs \\ \hline
Controlled-$H$                  & 6 single-qubit gates, 1 CNOTs \\ \hline
Controlled-$U(\theta,\phi,\delta)$ & 4 single-qubit gates, 2 CNOTs \\ \hline
\end{tabular}
}
\end{table}

\subsection{Custom Entanglement Gate Implementations}
\label{custom_gate}
Table \ref{table:2_qubit_gates} shows Qiskit's gate decomposition of some of the available two-qubit gates on IBM Quantum devices. We notice that CNOT is a base of these decompositions since it is a basis gate for IBM backends. Our main design principle is that PQCs do not necessarily need such standard two-qubit gates, but the goal is to use two-qubit entangling gates in general \cite{kandala}. Therefore, we utilize pulse-level access to quantum systems to design a faster entangling gate.

Figures~\ref{fig:cr_cnot}(c) and~\ref{fig:cr_cnot}(d) show the pulse schedule of two custom entanglement gates: $CR(\pi/4)$ and $CR(150ns)$ respectively. $CR(\pi/4)$'s implementation is based on the standard CNOT shown in Fig.~\ref{fig:cr_cnot}(b), where each pulse's \texttt{amp} and \texttt{duration} are carefully calibrated for each qubit. For example, the CNOT's CR tone directions intentionally avoid qubit-qubit collisions: accidentally driving terms with control-spectators \cite{magesan_cr}. Moreover, its pulses are calibrated to the largest amplitude without noticeable leakage in order to reduce the duration which the qubit is subject to decoherence~\cite{pranav_op, krantz_guide}. Thus, as its name suggests, $CR(\pi/4)$ uses the first CR tone in the CNOT's echoed cross-resonance sequence to achieve a $ZX$ rotation of $\pi/4$ (and uncancelled single-qubit rotations). We chose a fixed duration of $150$ns for our second entangling gate $CR(150ns)$ (similar to the gates used in~\cite{kandala}), with the goal of minimizing the effect of decoherence without compromising the optimization accuracy. The fixed duration is the average duration for the $CR(\pi/4)$ gate for different backend pairs. For the rest of pulse parameters in both CR gates, we chose to utilize the daily calibrations performed on IBM quantum devices. In this paper, we demonstrate how straightforward customized pulse gates, along with PQC analysis for parameters such as trainability, can lead to better algorithm performance. In future work, will explore optimizing the CR pulse parameters with different ansatze and study their correlation with trainability more closely. 

We generally modify the standard CNOT implementation in the following way:
\begin{itemize}
\item We removed the echoed CR sequence and replaced it with one, bare CR tone~\cite{sheldon_cr}. As a result, we also remove the $X$ rotations on the control channel associated with the sequence. This design choice was made for two reasons: First, by reducing the duration to less than half that of the basis CNOT, we sped up our custom gate compared to any standard two-qubit gate. With limited coherence times in today's noisy quantum systems, even small improvements to single- and two-qubit gate speeds are essential and can significantly enhance performance~\cite{qv_64}. Second, the echoed CR sequence (as mentioned in Section \ref{cr_dissect}) cancels uncontrolled single-qubit rotation terms such as $IX$ and $ZI$ in the CR Hamiltonian. We make the case that such terms are unwanted when the target unitary is CNOT, but not in our case.
\item We removed the target rotary pulses (target qubit pulses). As mentioned in Section \ref{cr_dissect}, these rotary echoes are used to suppress the driven $ZZ$ interaction and entanglements with target spectators~\cite{takita_2017}. We argue that entanglements with the target's spectators are not necessarily detrimental to the VQAs and this should be further explored.
\end{itemize}

In Section~\ref{results}, we use Hamiltonian Tomography (HT) to extract unitary representations of our custom gates, assuming a block-diagonal cross resonance Hamiltonian.
\begin{figure}[!t]
\vspace{0.2in}
  \begin{center}
  \includegraphics[width=3in, keepaspectratio]{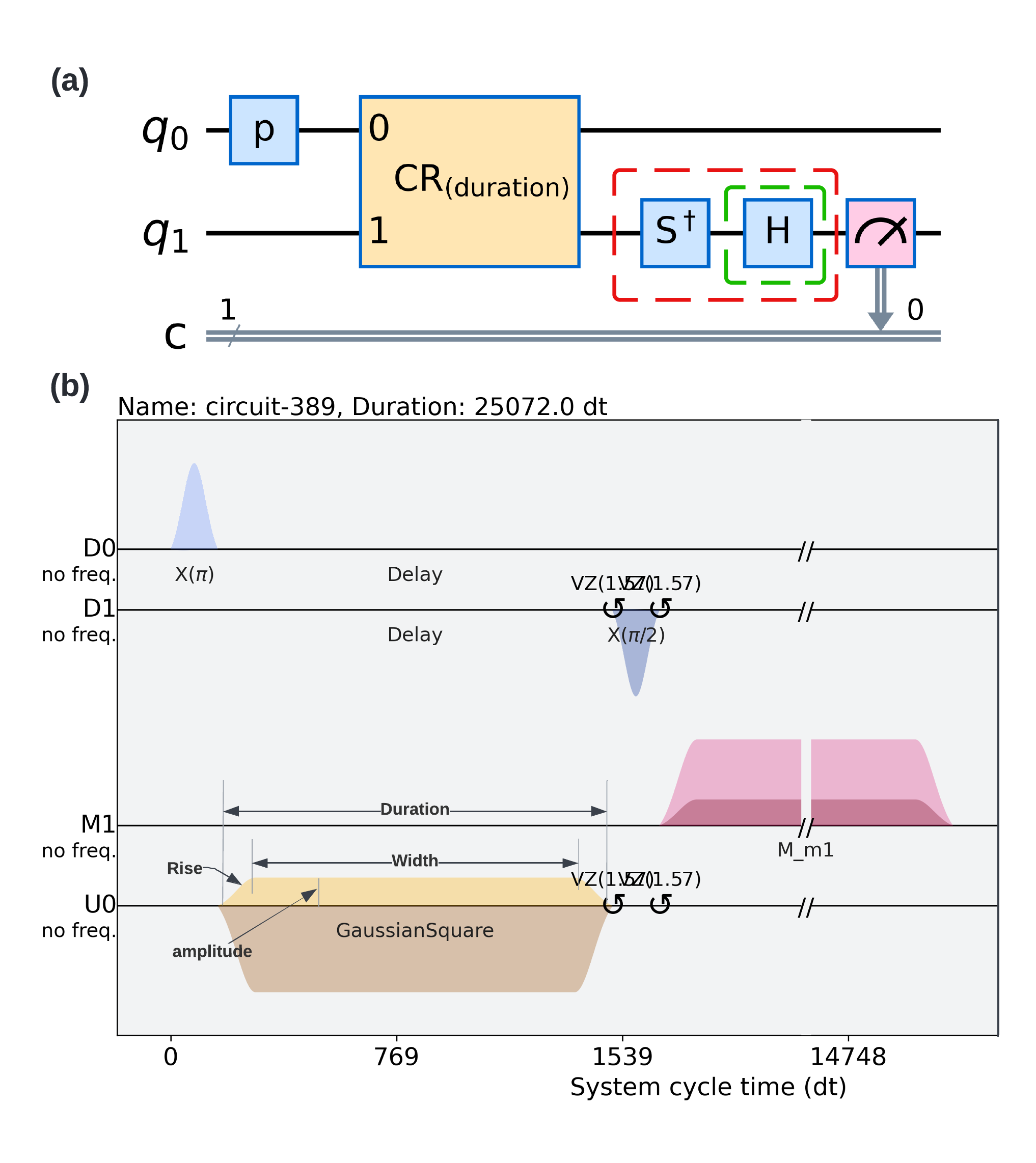} \\
  \caption{\textbf{(a)} The circuit diagram for the CR Hamiltonian Tomography experiment. The \texttt{p} gate is used to drive the control qubit to the 0 or 1 states. The target qubit is then measured by projecting it to the $X$ (green dotted square), $Y$ (red dotted square), and $Z$ bases. \textbf{(b)} Shows the pulse sequence for the experiment shown in (a) measuring the target qubit in the $X$ basis. The figure also shows the different pulse parameters changeable as part of the experiement.}
  \label{fig:ht}
  \end{center}
\vspace{-1em}
\end{figure}
\subsection{Characterizing CR-based Gates}
\label{cr_benchmarking}
Characterizing the pulse gates is essential to understanding their components and performance. For this purpose, we used Hamiltonian Tomography (HT), an accurate Hamiltonian calibration technique developed by Sheldon \textit{et al.} \cite{sheldon_cr}, to estimate the coefficients (strengths) $\omega$(s) of the CR Hamiltonian terms in (\ref{eqn:cr_ham}). 

Fig.~\ref{fig:ht}(a) and (b) show the gate and pulse sequence for HT. The experiment is performed by applying a CR tone (or the echoed CR tones for the standard implementation) with different durations. The target qubit is then measured by projecting to the $X$, $Y$, and $Z$ bases, with the control qubit either in the $0$ or $1$ state. The measurements (from the resulting six sets of experiments) are of the expectation values of each term in the Hamiltonian. It is important to note that HT is not sensitive to the $ZI$ term arising from a Stark shift (an off-resonant drive that dressed the qubit frequency) because the control qubit is in an eigenstate of the $Z$ operator. Thus, an additional Ramsey experiment on the control qubit was performed to estimate the strength of this term.
\begin{figure}[!tb]
    \centering
    \includegraphics[width=3in]{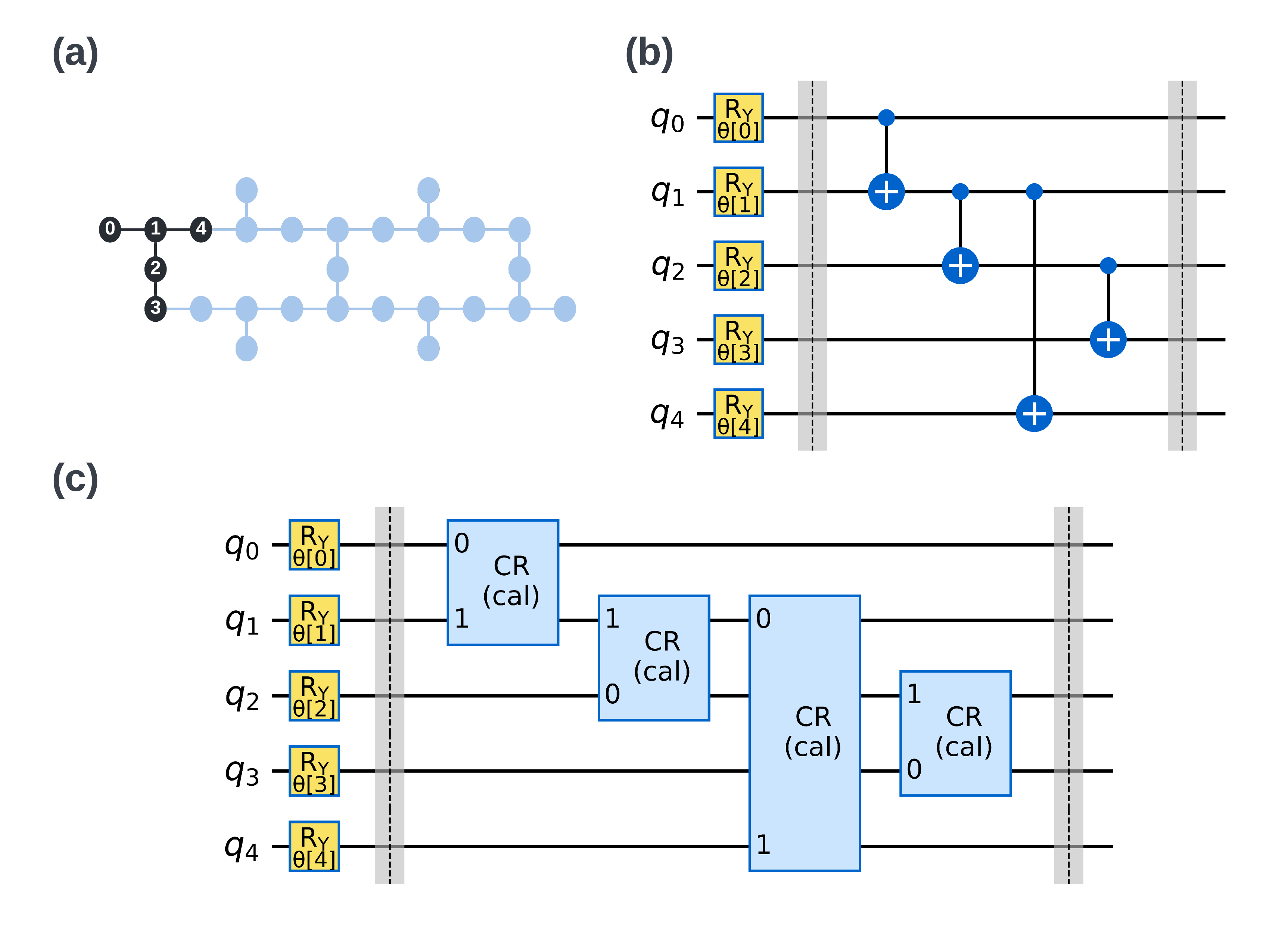}
    \caption{\textbf{(a)} The mapping of the two PQC configurations shown in (b) and (c) on \textit{ibmq\_montreal}. \textbf{(b)} The \textit{base} configuration using CNOT as entangling gates. \textbf{(c)} The \textit{CP\_ang} and \textit{CP\_dur} configurations using $CR(\frac{\pi}{4})$ or $CR(150ns)$ as entangling gates respectively.}
    \label{fig:pqcs}
    \vspace{-1em}
\end{figure}
Estimating the CR Hamiltonian terms can also be used to extract the unitary representation of custom gate implementations according to Schrodinger's equation
\begin{equation}
\label{eqn:schrodinger}
U_{H_{\rm CR}} = e^{-i H_{\rm CR}t} 
\end{equation}
where $H_{CR}$ is our CR Hamiltonian, and $t$ represents the CR tone's \texttt{duration}. This unitary can then be used to further analyze the PQCs for algorithmic descriptors such as expressibility, trainability, and entanglement.
% \begin{figure}[!t]
% \vspace{0.2in}
%   \begin{center}
%   \includegraphics[width=3in, keepaspectratio]{figures/cct_durations.png} \\
%   \caption{Execution time for different PQC configurations.}
%   \label{fig:duration}
%   \end{center}
% \vspace{-1em}
% \end{figure}
\section{Results and Evaluation}
\label{results}
\subsection{Experimental Setup}
We conducted our experiments on \textit{ibmq\_montreal}, a $27$-qubit backend available through IBM Quantum Services. The backend has average $T_1$ and $T_2$ times of $84.24$~$\mu$s and $85.32$~$\mu$s respectively, and an average CNOT error rate of $4.703\mathrm{e}{-2}$. Note that these values fluctuate and are monitored through daily calibrations available through Qiskit. We utilize Qiskit Runtime~\cite{runtime}, a programming model that allows for faster execution of quantum workloads on the cloud, to run our algorithm benchmarks.

Fig. \ref{fig:pqcs}(b) shows one layer of a base $n$-qubit PQC design. We will refer to PQCs constructed using this layer as \textit{base} PQCs. Fig \ref{fig:pqcs}(c) shows a single layer design utilizing our CR-based entanglers. We refer to PQCs utilizing these gates as \textit{Customized Pulse} PQCs or \textit{CP}. We refer to PQCs that use the $CR(\pi/4)$ gate as \textit{CP\_ang}, and PQCs utilizing $CR(150ns)$ as \textit{CP\_dur}. We used a linear entanglement arrangement in both circuits, which applies two-qubit gates to neighboring qubits only. The mapping of the circuits on \textit{ibmq\_montreal}'s topology is shown in Fig.~\ref{fig:pqcs}(a). 

In this section, we analyze the three PQC designs' circuit duration, expressibility, trainability, and entanglement. Next, we evaluate their performance for a set of chemistry and MaxCut problems.
\subsection{Circuit Duration}
We analyzed the three PQC configurations for total gate count, circuit depth, and duration. Since the three configurations share the same structure, they have identical gate counts and circuit depth (not shown). This was expected as our method only changes the pulse implementation of the entanglers and does not change the circuits' structure.

To measure the duration of \textit{base}, we compiled the circuit with the three levels of optimization available in Qiskit and picked the lowest duration. We left the measurement operation out of our speed calculations. As we mentioned in Section-\ref{custom_gate}, optimizations leading to faster quantum circuits are crucial as we are still competing with limited qubit coherence times. It is also critical as it gives more freedom to perform measurement pulses. 

We observe a speedup of up to $2.28\times$ in the execution time of \textit{CP\_ang} compared \textit{base}, with an average speedup of $(2.2\times)$. \textit{CP\_dur} on the other hand observes a maximum speedup of $(2.9\times$ over \textit{base}, with an average of $(2.8\times)$. This is a direct result of using faster two-qubit entangling gates. As shown in Figures \ref{fig:cr_cnot}(c) and \ref{fig:cr_cnot}(d), the custom gates are at least $(2\times)$ faster compared to standard CNOT. This reduction in duration is essentially equivalent to reducing the number of layers by half, as the echo pulse and subsequent $CR(\pi/4)$ are removed. As the $CR(150ns)$ use a fixed duration compared to $CR(\pi/4)$'s calibrated duration, \textit{CP\_dur} observes an average speedup of $(1.28\times)$ over \textit{CP\_ang}.
\subsection{CR Gates Characterization}
\label{ht_results}
Table~\ref{table:ht_terms} shows the characterization of the CR tones using HT. As expected, the CR tones have a higher strength of the $ZX$ entangling term compared to other Hamiltonian terms (except for $ZI$). As our CR implementations do not use an echoed pulse implementation, we see a high frequency for the $ZI$ term. However, this doesn't affect the algorithm performance as VQAs are unaffected by coherent terms which can be dealt with by the optimizer. Notably, the CR tones also experience a high frequency of the $IX$ term as a result of not using the echoed CR sequence.

As mentioned in Section~\ref{cr_benchmarking}, this characterization was used to obtain the unitary representation of our pulse gates according to~(\ref{eqn:schrodinger}) by substituting $t$ with the appropriate pulse duration (i.e. 150ns for $CR(150ns)$ and the calibrated duration for $CR(\pi/4)$). The unitaries were then used to analyze the PQCs in the following sections for expressibility, entanglement, and trainability.
\begin{table}[!t]
\centering
\caption{Average strength of CR Hamiltonian terms across all backend pairs}
\label{table:ht_terms}
\begin{tabular}{|c|c|}
\hline
\textbf{Term} & \textbf{Avg Frequency (MHz)} \\ \hline
\textbf{$\omega_{zx}$} & 0.69645487 \\ \hline
\textbf{$\omega_{zy}$} & -0.0112463 \\ \hline
\textbf{$\omega_{zz}$} & -0.04056 \\ \hline
\textbf{$\omega_{ix}$} & -0.1102794 \\ \hline
\textbf{$\omega_{iy}$} & 0.03167672 \\ \hline
\textbf{$\omega_{iz}$} & 0.03557382 \\ \hline
\textbf{$\omega_{zi}$} & 14.5783 \\ \hline
\end{tabular}%
\end{table}
\begin{figure}[!t]
  \begin{center}
  \includegraphics[width=\linewidth, keepaspectratio]{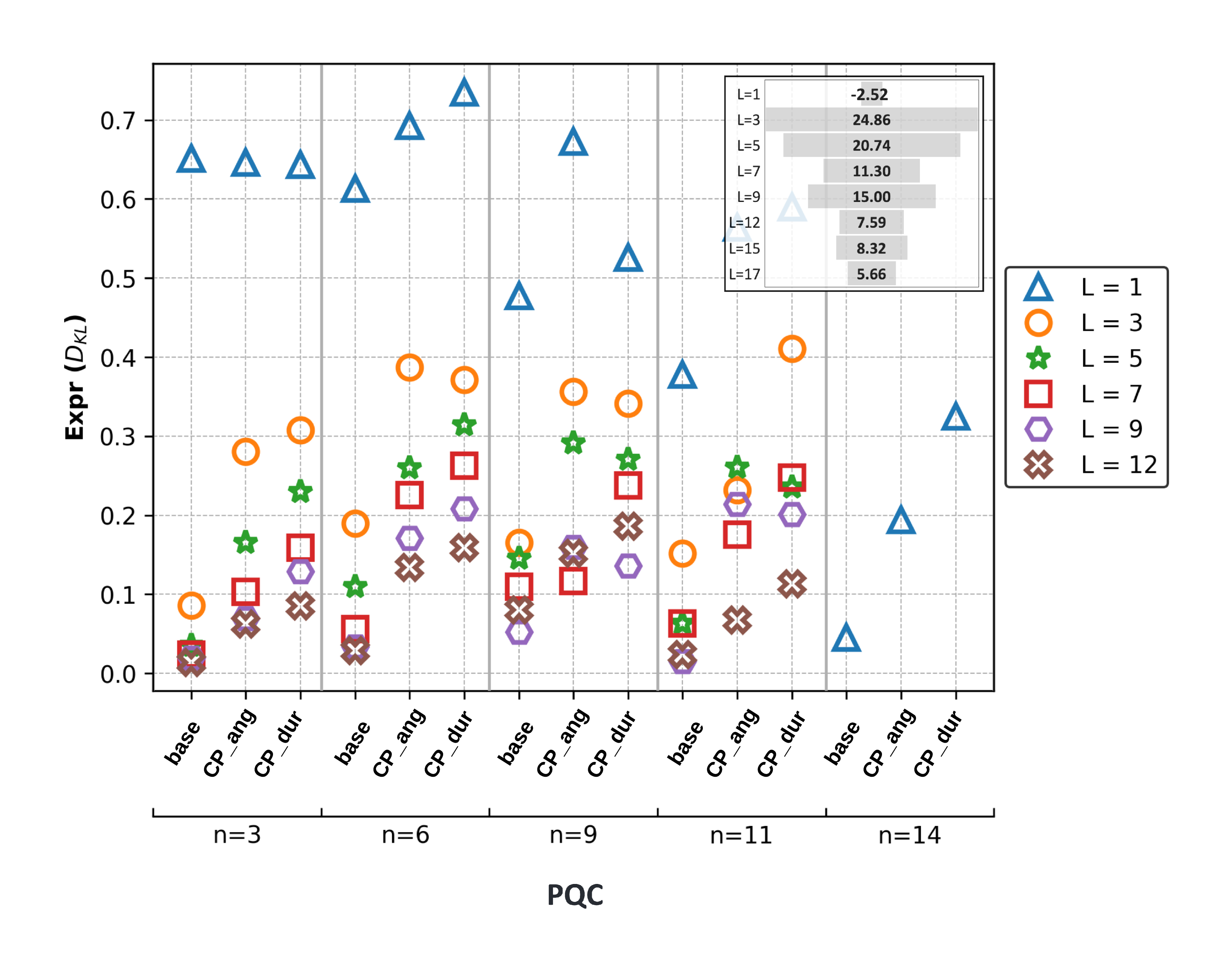} \\
  \caption{Expressibility of different PQC configurations ($n$ is number of qubits \& $L$ is the number of layers). \textbf{(Inset)} Shows the average increase in expressibility for \textit{base} over \textit{CP} PQCs as a function of the number of layers.}
  \label{fig:expr}
  \end{center}
\vspace{-1em}
\end{figure}
\begin{figure}[!t]
  \begin{center}
  \includegraphics[width=0.9\linewidth, keepaspectratio]{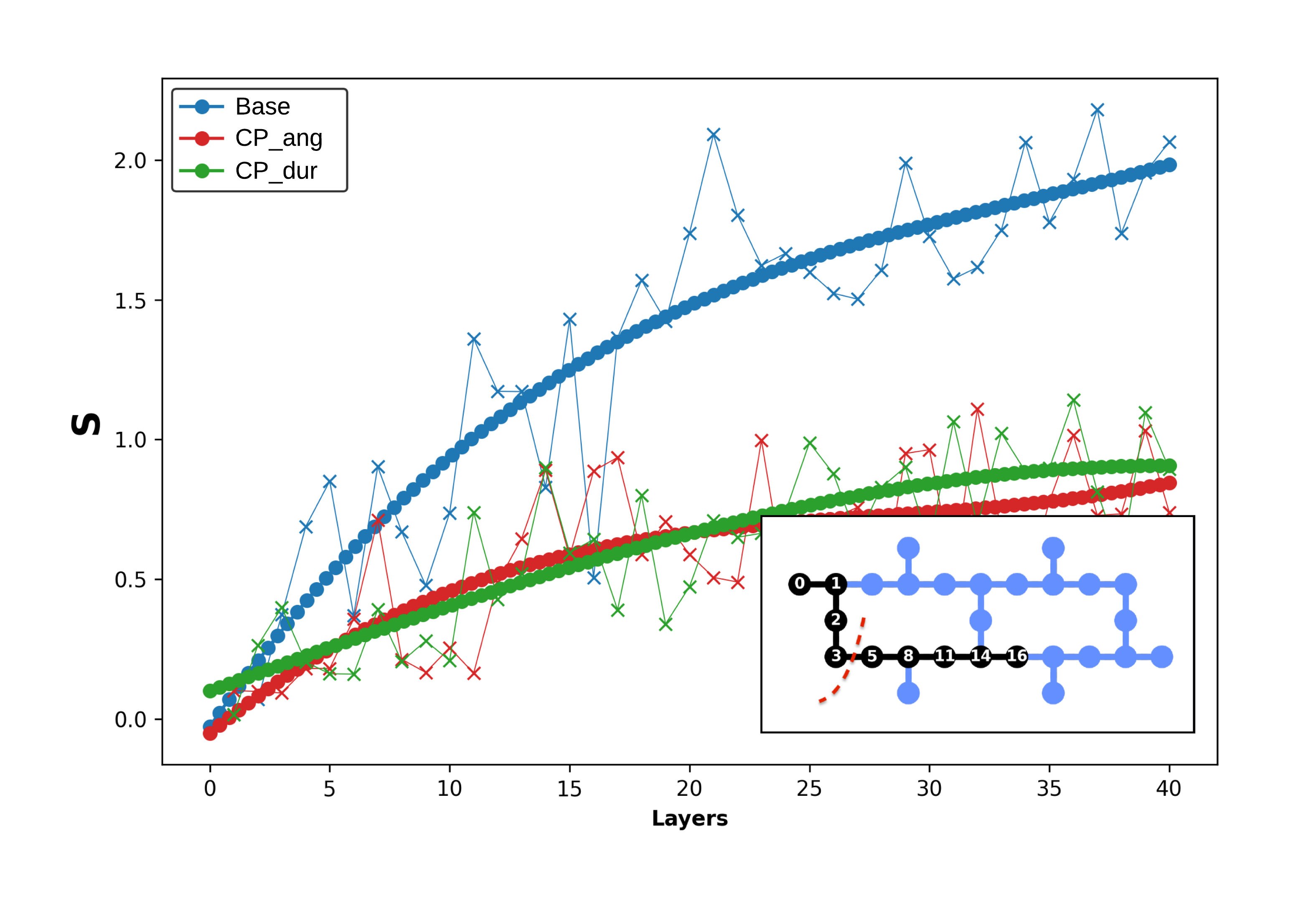} \\
  \caption{The trend lines of entanglement entropy $S$ for the three PQC configurations vs. circuit depth (Layers) for a $4$-$5$ qubit bipartition, as defined in~(\ref{eqn:entropy}) and illustrated in the \textbf{inset} of the figure.}
  \label{fig:entang}
  \end{center}
\vspace{-2em}
\end{figure}
\subsection{Expressibility}
We used state-vector simulation to perform the necessary sampling for expressibility calculation, as detailed in Section-\ref{expr_entang}. As we mentioned in Section~\ref{ht_results}, the unitary representations of the CR-based gates were used in the sampling of the \textit{CP} PQCs.

Fig.~\ref{fig:expr} shows the expressibility of the the three PQC configurations with varying numbers of qubits and layers. As mentioned in the background, a lower value means better expressibility for the circuit using the KL divergence measure. For PQCs with a number of layer ($L>1$), we see that the \textit{base} configuration is more expressive than \textit{CP}. The inset of Fig.~\ref{fig:expr} shows the average increase in expressibility of \textit{base} over \textit{CP} as a function of $L$. \textit{base} observe a higher average increase in expressibility with shallow numbers of layers, with a maximum of $24\%$ over \textit{CP} PQCS at $L=3$. The difference in expressibility gradually decreases as we add more layers and expressibility values saturate.

This reduction in the \textit{CP} PQCs' expressibility is not exactly harmful to the performance. As we mentioned in Section~\ref{expr_entang}, findings from Holmes \textit{et al.} \cite{holmes_expr} indicated that the more expressive the PQC, the smaller the variance in cost gradients and hence, the harder it is to train. Their results also suggest that ansatze need not be highly expressive; instead, it is more important that they are trainable and contain a solution to the problem. With that, the \textit{CP} configuration proves to be more trainable (Section~\ref{train_results}). Our algorithm performance results (Section~\ref{algo_results}) further confirm that this reduction in expressibility does not harm the algorithm performance and can, in fact, optimize it.
\subsection{Entanglement}
Fig. \ref{fig:entang} shows the trend of entanglement entropy for the three PQC configurations with increasing circuit depth. The results are obtained for $9$-qubit PQCs with a $4$-$5$ partition, as shown in the figure. We see from the trend lines that \textit{base} always creates more entanglement compared to \textit{CP} PQCs. The \textit{base} configuration has an entropy that is, on average, $2.58\times$ higher than \textit{CP}'s across all circuit depths. Such reduction in entanglement is expected due to the short durations of the CR tones used in \textit{CP} compared to CNOT. We also see that the entropy difference drops as we increase the circuit depth before the values reach saturation.

In regards to capturing the PQCs entanglement more accurately, this can be further improved by accounting for spectator entanglements as well. As mentioned in Section~\ref{cr_dissect}, the CR interaction on transmons can also generate coherent terms due to coupling with the target's nearest-neighbors or spectators. As we also mentioned, the target rotaries in the echoed CR sequence $ECR$ are proven to suppress this type of entanglements \cite{nereeja_cr}. Thus, our $CR(\frac{\pi}{4})$ pulse can possibly have more spectator entanglements, which can lead to entirely different entanglement dynamics. Accounting for spectator interactions, however, requires additional experimentation. This can be done by using generic quantum tomography techniques or, more favorably, the Hamiltonian Error Amplifying Tomography (HEAT) technique proposed by Sundaresan \textit{et al.}~\cite{nereeja_cr}.

\begin{figure}[!htp]
\centering
\includegraphics[width=\columnwidth]{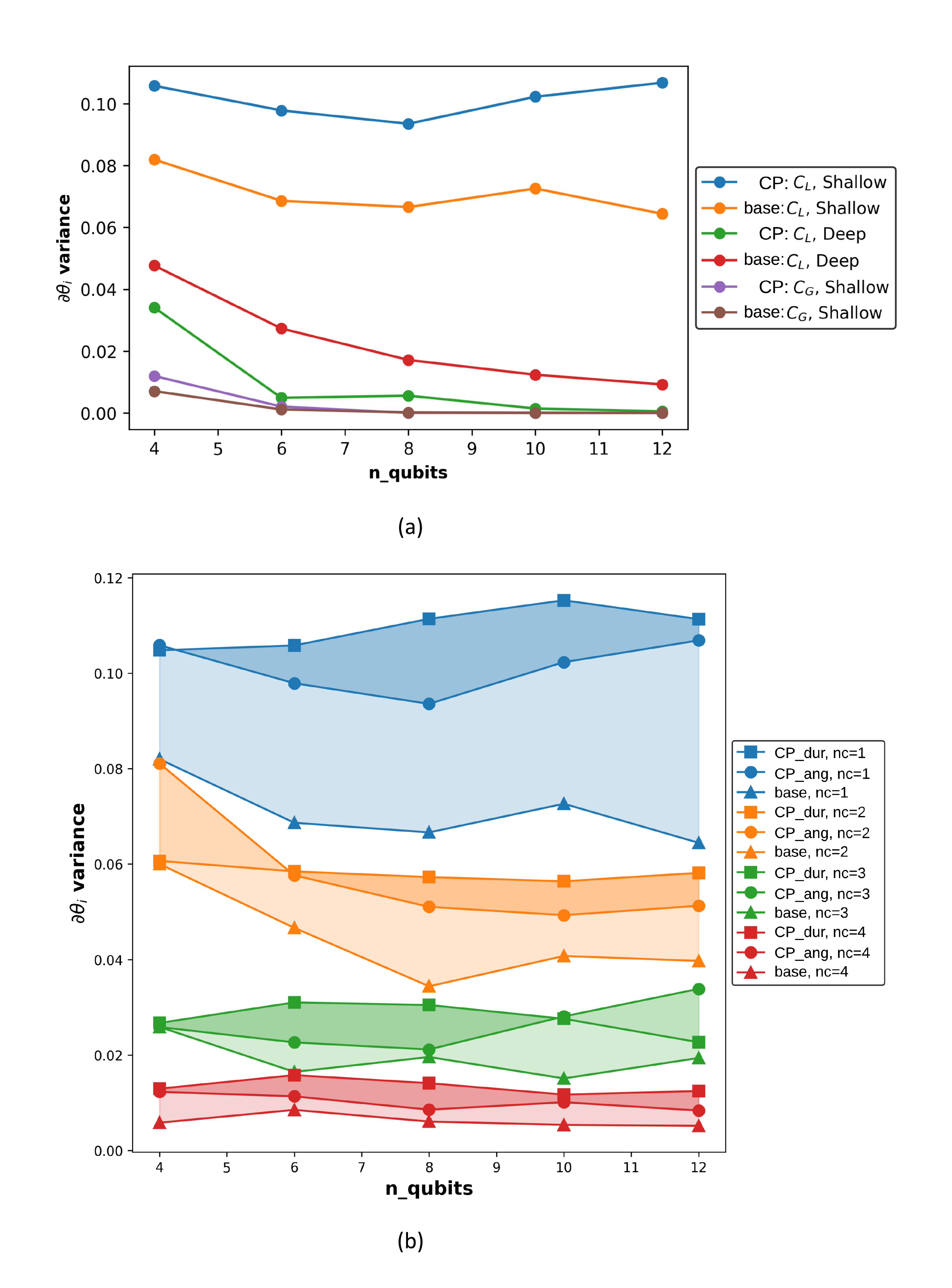}%
\caption{\textbf{(a)} The change in variance of the partial cost function derivative for \textit{base} and \textit{CP\_ang} vs. PQC size. Shallow PQCs use a number of layers $L=\log_2(N)$ where $N$ is the number of PQC qubits, while Deep PQCs use a polynomial number ($L=10*N$). The local cost functions shown here are for $N_C = 1$. \textbf{(b)} The variance vs. circuit size for shallow PQCs and different values of $N_C$. The shaded regions highlight the difference between \textit{CP\_dur}, \textit{CP\_ang}, and \textit{base} configurations.}
\label{fig:train}
\end{figure}

\begin{figure}[!t]
  \begin{center}
  \includegraphics[width=\linewidth, keepaspectratio]{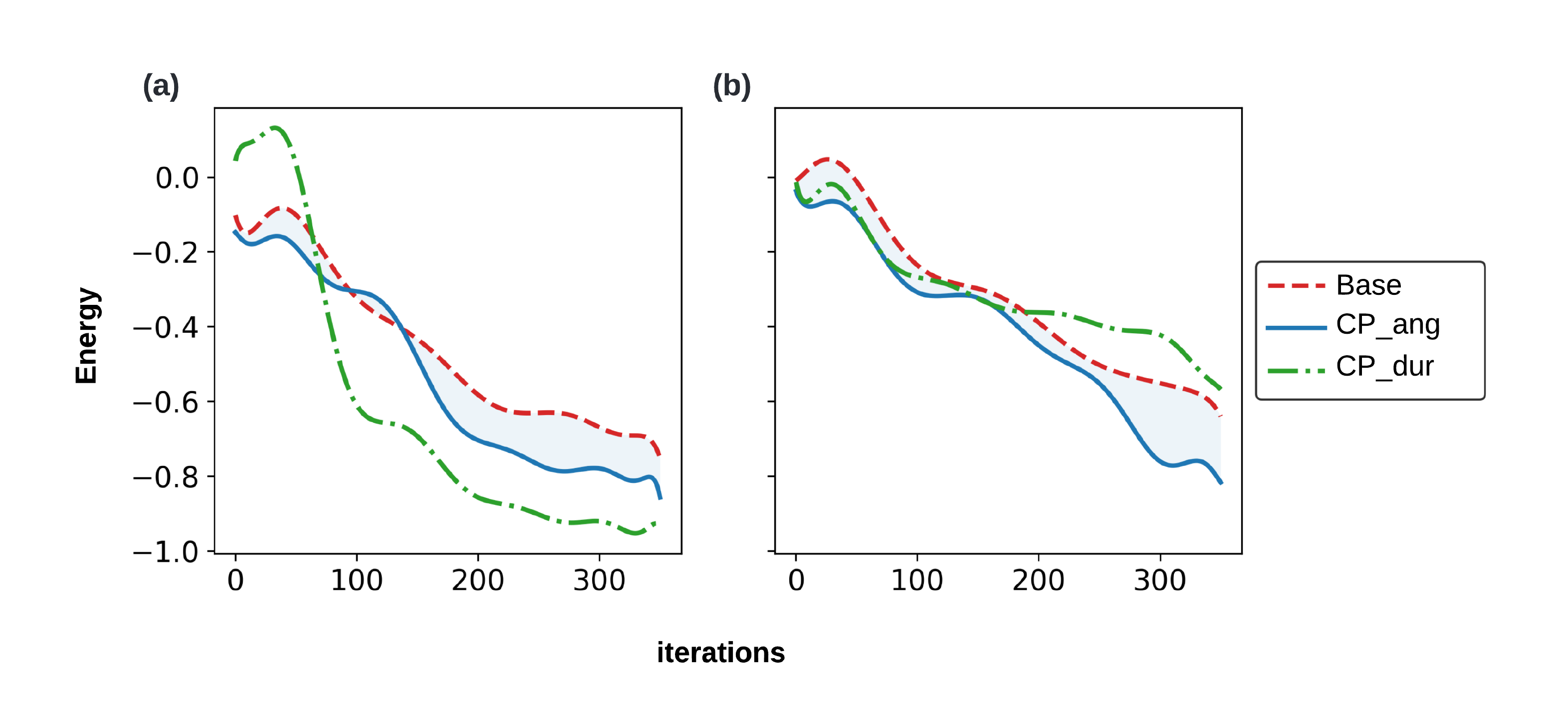} \\
  \caption{VQE performance of the three PQC configurations for two Hamiltonian mappings: \textbf{(a)} Bravyi-Kitaev (BK) and \textbf{(b)} Jordan-Wigner (JW). All runs used the Simultaneous Perturbation Stochastic Approximation (SPSA) \cite{spsa} gradient-based optimizer.}
  \label{fig:jw_vs_bk}
  \end{center}
\vspace{-1em}
\end{figure}
% \begin{figure*}[tb]
% \centering
% \subfloat[\label{fig:H2}]
%   {\includegraphics[width=.33\textwidth]{figures/H2.png}}\hfill
% \subfloat[\label{fig:LiH}]
%   {\includegraphics[width=.33\textwidth]{figures/LiH.png}}\hfill
% \subfloat[\label{fig:BeH2}]
%   {\includegraphics[width=.33\textwidth]{figures/BeH2.png}}
% \caption{VQE results for \textbf{(a)} H$_2$, \textbf{(b)} LiH, and \textbf{(c)} BeH$_2$ molecules for each PQC configuration. The Hamiltonian settings for each molecule is shown in Table~\ref{table:chem_hams}. Note that \#iterations here corresponds to the total number of objective function evaluations and not the number of SPSA opimization iterations.}
% \label{fig:chem}
% \end{figure*}
\begin{figure*}
    \centering
    \includegraphics[width=\textwidth]{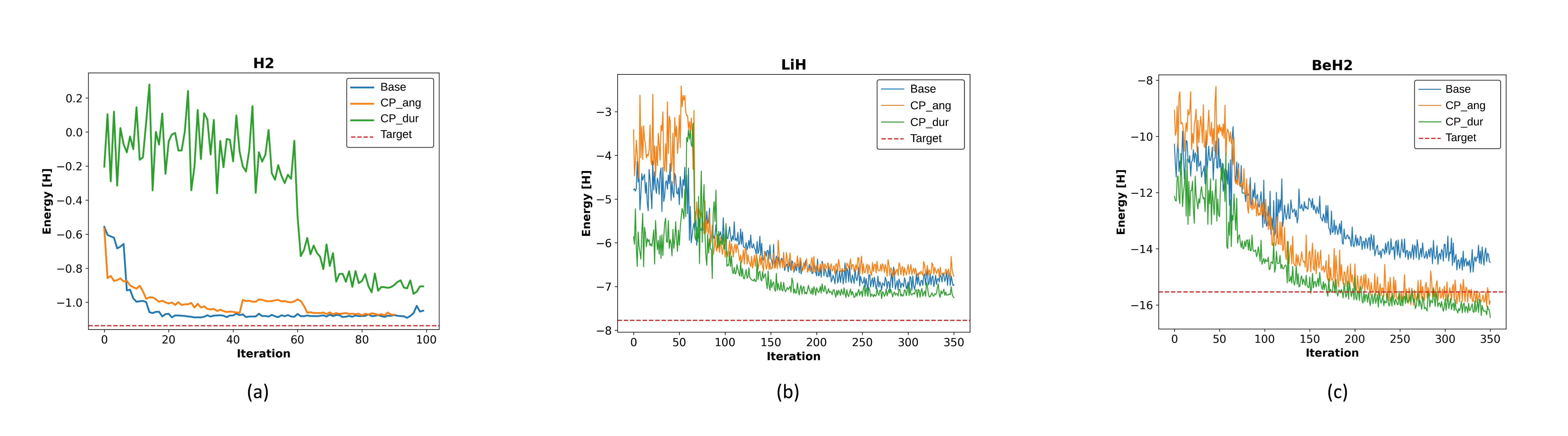}
    \caption{VQE results for \textbf{(a)} H$_2$, \textbf{(b)} LiH, and \textbf{(c)} BeH$_2$ molecules for each PQC configuration. The Hamiltonian settings for each molecule is shown in Table~\ref{table:chem_hams}. Note that \#iterations here corresponds to the total number of objective function evaluations and not the number of SPSA opimization iterations.}
    \label{fig:chem}
\end{figure*}
\subsection{Trainability}
\label{train_results}
To analyze the PQC configurations' trainability, we follow a cost-function-based analysis similar to that in \cite{cerezo_bp} and \cite{patti_bp}. We used a simple ground state preparation problem, which can be defined by the global cost function
\begin{equation}
C_G = 1 - p_{\ket{0}^{\otimes N}}
\end{equation}
where $N$ is the total number of qubits, and $p_{\ket{0}^{\otimes N}}$ is the probability of measuring the $\ket{00...0}_N$ state. For the local cost function, we only consider the probability of a subset of qubits
\begin{equation}
C_L = 1 - p_{\ket{0}^{\otimes N_C}}
\end{equation}
where $N_C$ is the number of cost-function qubits. It is interesting to point out that for ($N_C = 1$), $C_L$ has a cost landscape similar to that of a local cost function acting on each qubit separately \cite{pennylane_bp}.

Fig.~\ref{fig:train}(a) shows the results for different cost function and PQC settings. The bottom two lines (Deep, $C_G$) follow the conclusions from \cite{cerezo_bp} that this cost function, and others like it, exhibit barren plateaus. The figure also proves that local cost functions like $C_L$ will exhibit barren plateaus for deep numbers of layers. We see that \textit{base} observes better variance for small numbers of qubits, but both curves are exponentially decreasing due to barren plateaus. More interestingly, we see that \text{CP\_ang} has better local cost function trainability with shallow layers. This observation is further expanded in Fig.~\ref{fig:train}(b) for the three PQC configurations, which confirms it for different values of $N_C$ (up to a certain limit). The results also suggest that the performance gap (indicated by the shaded regions) shrinks with increasing $N_C$. This better overall local cost function trainability can be attributed to the CP PQC's reduced expressibility, entanglement, and duration; as each of these parameters is proven to negatively affect training \cite{holmes_expr, patti_bp, noise_bp}.

To further explore the advantages of local cost function training, we compare the VQE optimization performance of two different $4$-qubit Hamiltonians for the H$_2$ molecule. The Hamiltonians $(H_{\rm JW}, H_{\rm BK})$ were obtained using two of the most commonly used techniques to map fermionic to spin operators: Jordan-Wigner (JW) \cite{mike_n_ike} and Bravyi-Kitaev (BK) \cite{parity}. As discussed in \cite{cerezo_bp, cerezo_extra}, BK mapping often leads to more local Pauli terms and hence to more trainable cost functions. 

Fig.~\ref{fig:jw_vs_bk} shows the results from the experiment ran on \textit{ibmq\_montreal}. Contrary to our expectations based on the local cost function analysis, \textit{CP\_dur} performs poorly with the BK mapping compared to other PQC configurations and its JW performance. This is an important finding as it reveals that other factors (yet to be determined) besides the locality of the cost function affect \textit{CP} performance. On the other hand, we see a larger performance gap between \textit{CP\_ang} and \textit{base} for BK compared to JW mapping (the shaded areas in the figure), indicating that the performance was affected by the locality of the Hamiltonian. Although this somewhat confirms the results from Fig.~\ref{fig:train} showing that \textit{CP} has better local cost function trainability than \textit{base}, the two Hamiltonian mappings had similar performances for each PQC configuration. Additionally, both mappings fall short in terms of performance compared to an H$_2$ mapping that uses $2$-qubits (Section~\ref{algo_results}). Overall, we believe that utilizing efficient local cost function implementations can lead to better performance (as proven in \cite{cerezo_extra}) using pulse-optimized gates, and we leave this exploration for future work.
\begin{figure*}[!tb]
\centering
  \begin{center}
  \includegraphics[width=\textwidth]{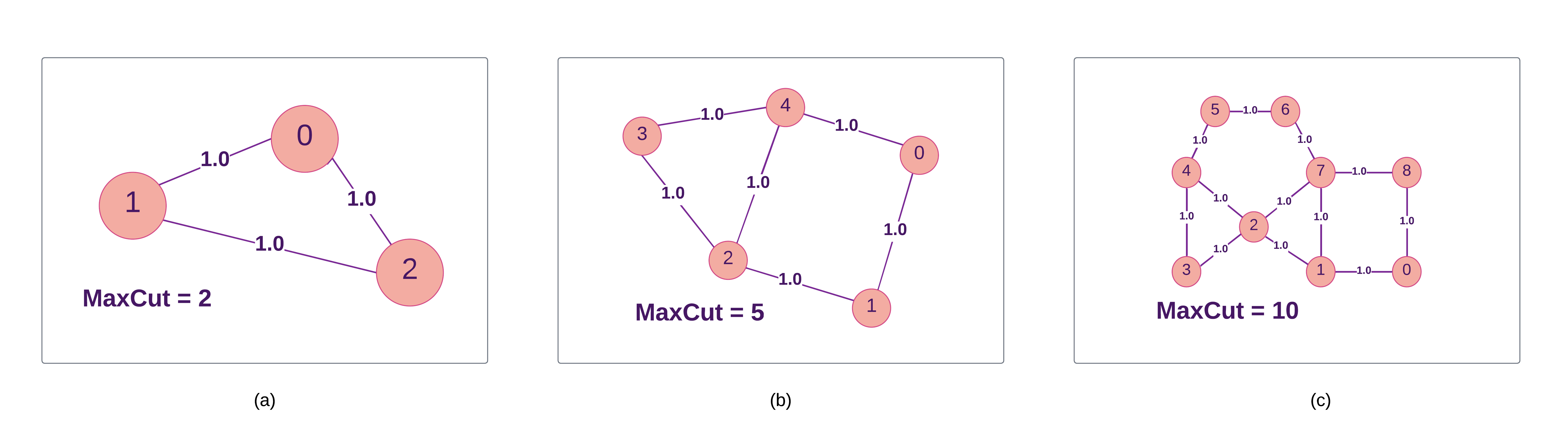} \\
  \caption{The $3$-, $5$-, and $9$-node graphs used for MaxCut are shown in \textbf{(a)}, \textbf{(b)}, and \textbf{(c)}, respectively, with their classically calculated MaxCut values.}
  \label{fig:maxcut}
  \end{center}
\end{figure*}
\subsection{Algorithm Performance}
\label{algo_results}
We compare the performance of the three PQC configurations with two sets of VQE applications from chemistry and optimization. In chemistry, we use VQE to find the ground state energy of the H$_2$, LiH, and BeH$_2$ molecules, which corresponds to finding the minimum eigenvalue of Hermitian matrices characterizing these molecules. For optimization, we solve three MaxCut problems (shown in Fig.~\ref{fig:maxcut}). We ran all our benchmarks on \textit{ibmq\_montreal} accessed through IBM Cloud and configured our experiment as follows. We used the Simultaneous Perturbation Stochastic Approximation (SPSA) \cite{spsa} as our optimization routine, with the maximum number of iterations set to $100$. We use an ($R_YR_Z$) rotation (instead of the $R_Y$ shown in Fig.~\ref{fig:pqcs}) for chemistry benchmarks. The number of layers was set to $5$ for both \textit{CP} and \textit{base} across all applications.

\subsubsection{Chemistry Benchmarks}
Fig.~\ref{fig:chem} shows VQE results for three chemistry molecules: H$_2$, LiH, and BeH$_2$. The Hamiltonians were obtained through Qiskit's integration with the PySCF library \cite{pyscf}. We favor reducing the number of qubits guided by a quick analysis of trainability for H$_2$, which revealed that H$_2$'s 2-qubit mapping has a variance in partial gradients $var[\frac{\partial C}{\partial \theta_o}]$ that is $3\times$ higher than that for the $4$-qubit mapping. Therefore, we chose Jordan-Wigner and Parity \cite{parity} mappings to map our molecules' fermionic operators to spin operators. The Parity mapping was chosen for the H$_2$ and LiH, as it allowed for reducing the number of qubits by utilizing $Z_2$ symmetries. Table~\ref{table:chem_hams} summarizes the experiments' configurations for each molecule.

Fig.~\ref{fig:chem}(a) shows the results for the H$_2$ molecule. Both \textit{base} and \textit{CP\_ang} PQCs fail to find the lowest energies, but their results are fairly and equally close to the exact solution while \textit{CP\_dur} performs slightly worse, which indicates a lower quality for this PQC with small configuration. For the LiH molecule shown in Fig.~\ref{fig:chem}(b), both \textit{base} and \textit{CP} PQCs results are fairly close to the exact solution, with \textit{CP\_dur} being slightly closest to the exact solution. For the $6$-qubit BeH$_2$ problem shown in Fig.~\ref{fig:chem}(c), \textit{CP} PQCs clearly outperform \textit{base}, with the lowest energy obtained through \textit{CP\_ang} reaching chemical accuracy (defined to be within $0.0016$ Hartree of the exact result). This result indicates that the \textit{CP} configurations may have better potential with larger and more complex problem structures.
\begin{table}[]
\caption{Hamiltonian configurations for chemistry molecules}
\label{table:chem_hams}
\centering
\resizebox{\columnwidth}{!}{%
\begin{tabular}{|c|c|c|c|}
\hline
\textbf{Molecule} & \textbf{Mapping} & \textbf{\begin{tabular}[c]{@{}c@{}}Interatomic Distance \\ (Angstrom)\end{tabular}} & \textbf{\# of Qubits} \\ \hline
\textbf{H$_2$}     & Parity           & 0.72                                                                                & 2                     \\ \hline
\textbf{LiH}      & Parity           & 2.5                                                                                 & 4                     \\ \hline
\textbf{BeH$_2$}     & Jordan-Wigner    & 1.5                                                                                 & 6                     \\ \hline
\end{tabular}%
}
\end{table}
\begin{table}[tb]
\caption{\# of correct MaxCut solutions and ROCA for the top-$5$ probabilities}
\label{table:result_maxcut}
\resizebox{\columnwidth}{!}{%
\begin{tabular}{c|ccc|ccc|}
\cline{2-7}
 & \multicolumn{3}{c|}{\textbf{\# of Correct Solutions}} & \multicolumn{3}{c|}{\textbf{ROCA}} \\ \hline
\multicolumn{1}{|c|}{\textbf{MaxCut Problem}} & \multicolumn{1}{c|}{\textbf{base}} & \multicolumn{1}{c|}{\textbf{CP\_ang}} & \textbf{CP\_dur} & \multicolumn{1}{c|}{\textbf{base}} & \multicolumn{1}{c|}{\textbf{CP\_ang}} & \textbf{CP\_dur} \\ \hline
\multicolumn{1}{|c|}{\textbf{3-nodes}} & \multicolumn{1}{c|}{4/5} & \multicolumn{1}{c|}{5/5} & 5/5 & \multicolumn{1}{c|}{1} & \multicolumn{1}{c|}{1} & 1\\ \hline
\multicolumn{1}{|c|}{\textbf{5-nodes}} & \multicolumn{1}{c|}{2/5} & \multicolumn{1}{c|}{2/5} & 2/5 & \multicolumn{1}{c|}{1} & \multicolumn{1}{c|}{1} & 1\\ \hline
\multicolumn{1}{|c|}{\textbf{9-nodes}} & \multicolumn{1}{c|}{0/5} & \multicolumn{1}{c|}{1/5} & 0/5 & \multicolumn{1}{c|}{0} & \multicolumn{1}{c|}{2} &  0\\ \hline
\end{tabular}%
}
\end{table}
\subsubsection{MaxCut}
For MaxCut benchmarks, the optimized set of parameters obtained by VQE was first used to prepare a quantum state through the PQC. This state was then sampled to construct an eigenstate, from which the highest probabilities correspond to MaxCut solutions (graph partitionings). The solutions can then be evaluated by calculating their cut values and comparing them to a classically calculated MaxCut reference. Table~\ref{table:result_maxcut} shows the number of correct Maxcut solutions out of the top$-5$ solutions for each PQC configuration. The table also shows results using the Rank of Correct Answer (ROCA) metric proposed by Tannu \textit{et al.}~\cite{tannu_roca}, which, as its name suggests, accounts for the order of appearance of the correct answer(s).

We see that \textit{CP\_ang} generally performs better than \textit{base} and \textit{CP\_dur}, specifically for the $3$- and $9$-node problems. This is evident for the $9$-node case, where \textit{CP\_ang} configuration was capable of finding the correct solution with a ROCA of 2 compared to 0 correct solutions for the two other configurations.

In conclusion, we see that the \textit{CP} configurations have, on average, a better algorithmic performance compared to \textit{base}, specifically for larger problem instances ($6$-qubit BeH$_2$ molecule and $9$-node MaxCut). The CP's reduction of expressibility, entanglement, and duration prove beneficial to the algorithm's performance and trainability. We also observe that \textit{CP\_ang} performs better than \textit{CP\_dur} in general, which shows the sensitivity of the algorithm's performance to tuning the CR pulse. We argue that optimizing pulse parameters, alongside utilization of efficient local cost-functions can lead to further improvements.
\section{Related Work}
The majority of work done to optimize PQCs focused on the higher levels of the algorithm, such as analyzing and improving their trainability of PQCs \cite{mcclean_bp, cerezo_bp, patti_bp, holmes_expr, ent_bp, noise_bp}, parameter initialization methods \cite{grant_init, meta_vqe}, and developing optimizers and optimization strategies that are tailored for VQAs \cite{avqe, evqe, q_gradient, layer_pqc, aavqe, sukin_pect, axis_vqe, gushu_isca, how_much_ent}.

Hardware-efficient PQCs have been first proposed by Kandala \textit{et al.}~\cite{kandala}. Their work used fixed-duration entanglers to simulate the performance of VQE for small molecules and quantum magnets. In this work, we extend their usage of Hamiltonian tomography by utilizing its data to analyze our PQCs for expressibility, trainability, and entanglement. Recent studies have explored hardware-oriented VQAs' analysis and optimization. Ravi \textit{et al.} \cite{vqe_ravi} proposed a VQA error-mitigation approach that tunes single qubit gate scheduling and dynamical decoupling sequences in a variational approach. Other works have also explored the effects of noise on VQAs and hardware-efficient PQCs~\cite{vqe_noise, vqe_noise_2, vqe_noise_3}. The work by~\cite{vqe_noise_3} determines optimal PQC depth at different noise levels and investigates the circuit resiliency to noise with the inclusion of redundant parameterized gates. Zeng \textit{et al.}~\cite{vqe_noise_2} simulates specific hardware-efficient PQCs' performance with different noise models and noise levels. Their results showed that VQE's performance degrades as the noise probability or the circuit depth increase. A more recent study by Saib \textit{et al.}~\cite{vqe_noise} discussed the effect of noise on chemistry applications and profiled various PQCs for expressibility. Their results suggest that expressibility is weakly correlated to VQE performance. We note that the original expressibility and entanglement paper by~\cite{sukin_expr} states that it has not yet discovered an accurate correlation between these measures and VQE applications. Our work aims to uncover ways to merge PQC descriptors, such as trainability, expressibility, and entanglement, to hardware-specific parameters in PQC design.

More recently, VQA optimization through Quantum Optimal Control (QOC) has gained more attraction \cite{pGrape, ctrlVQE, vqe_extra, vqe_extra_2}. For hardware-efficient gate-based PQCs, Liang \textit{et al.} \cite{vqe_pulse} proposed a pulse optimization framework that manipulates the PQC gate amplitudes as part of the VQA optimization routine. In contrast to their approach, we choose to preconfigure our CR pulse parameters and not attach them to the VQA optimization procedure, as it was proven in \cite{kandala} through numerical simulations that accurate optimizations can be obtained for fixed-phase two-qubit gates. Additionally, as over-parameterization of pulses can lead to difficulties in optimization~\cite{ctrlVQE}, our approach leads to a lower number of parameters and, ultimately, a faster VQA implementation as the circuit size grows. A more recent work by the same group~\cite{pan} proposes a progressive pulse-ansatz construction and learning approach utilizing non-gradient optimizers to generate more scalable and efficient anstaze.

Utilizing pulse access for faster two-qubit gate implementation has been explored by~\cite{nate_pulse,qv_64, pranav_op, pulse_scaling}. Jurcevic \textit{et al.}~\cite{qv_64} experimented with a \textit{direct CNOT} approach that uses compensation mechanisms different than the echoed CR implementation to achieve a higher quantum volume of $64$ on IBM machines. Gokhale \textit{et al.}~\cite{pranav_op} utilized OpenPulse and knowledge of the CR gate to implement a more efficient $R_{ZZ}$ rotation, which is a core operation for quantum chemistry and optimization algorithms. Their optimized implementation experienced both error rate and execution time reductions and has been adopted by Qiskit's transpiler, as shown in Table~\ref{table:2_qubit_gates}. More recently, Stenger \textit{et al.}~\cite{pulse_scaling} proposed a pulse-scaling method that scales the area of the CR and rotary pulses to create $R_{ZX}(\theta)$ rotations. Their method improves the gate fidelity with no additional calibrations. Their work has been further extended in~\cite{nate_pulse} to arbitrary gates and to develop a pulse-efficient circuit transpilation framework, which decomposes two-qubit gates into the hardware-native $R_{ZX}$ rather than the CNOT-based transpilation.
\section{Conclusions}
In this work, we utilize pulse-level access to quantum machines to alter the standard design of two-qubit gates. Additionally, we identify a suitable combination of hardware and algorithmic parameters that can be efficiently embedded in the design and are proven to impact performance. Our analysis results prove that our customized pulse implementations maintains similar expressibility to a standard PQC and is more trainable for local cost functions, all while reducing the circuit duration to half. Therefore, this implementation is more suitable for VQAs. Our algorithm performance results show that in at least three cases, our customized pulse PQC configuration outperforms the base implementation. As previous literature closely ties PQC parameters such as entanglement, noise, and expressibility to barren plateaus, we believe that pulse optimization, which directly impacts said parameters, is a very promising approach to enhance trainability. We leave this as our main future goal. Other next steps include designing a comprehensive entanglement model of the PQC by including spectator entanglements and further testing with a more diverse set of PQC architectures.

\section{Acknowledgement}
M.I. would would like to thank the NSF QISE-NET Fellowship for funding through the grant DMR 17-47426, and the IBM Quantum Hub at NC State for access to \textit{ibmq\_montreal}.
\bibliographystyle{IEEEtran}
\bibliography{references}
\end{document}